\documentclass[twocolumn,showpacs,preprintnumbers,amsmath,amssymb, prl]{revtex4}

\usepackage{graphicx}
\usepackage{dcolumn}
\usepackage{bm}
\usepackage{epsfig}

\begin{document}

\title{Computer Simulation on Terahertz Emission from Intrinsic Josephson Junctions of
High-$T_c$ Superconductors}

\author{Shizeng Lin\(^{1,3}\), Xiao Hu\(^{1}\) and Masashi
Tachiki\(^{2}\)}

\affiliation{ \(^{1}\) National Institute for Materials Science,
Tsukuba 305-0047, Japan
\\ \(^{2}\)Graduate School of Frontier Sciences, The University of Tokyo,
Kashiwanoha 277-8568, Japan
\\ \(^{3}\)Zhejiang Institute of Modern Physics, Zhejiang University,
Hangzhou 310027, P.R. China}
\date{May 21, 2007}

\begin{abstract}
Solving coupled nonlinear sine-Gordon equations and Maxwell
equations numerically, we study the electromagnetic and
superconducting properties of the single crystal of high-$T_c$
superconductor $\rm{Bi_2Sr_2CaCu_2O_{8+\delta}}$ with a static
magnetic field applied parallel to the $ab$-plane and a dc current
fed in along the $c$-axis. Cavity resonances of transverse plasma
occur in the intrinsic Josephson junctions with frequencies in
terahertz regime. It is revealed that the electromagnetic wave can
transmit from the junctions into space. The emitted energy counted
by the Poynting vector is about $400\rm{W/cm^2}$. The frequency as
well as the energy of emission can be tuned almost continuously by
the current and magnetic field.

\end{abstract}


\maketitle

\noindent {\it Introduction --} Terahertz (THz) technology is an
extremely attractive field. The main users of the THz
electromagnetic waves are perhaps the biomedical diagnostics, DNA
probe and cancer detection; a THz tomography is also very useful for
material characterization\cite{xczhang02}. Although the lower and
higher frequency bands of electromagnetic field can be generated by
electronics and photonics respectively, seeking solid-state, stable
generators for the THz waves is still a subject of scientific
effort. While recently the semiconductor heterostructures generate
THz emissions with high efficiency\cite{Kohler}, known as the
quantum cascade lasers, the issue of frequency tunability has not
been resolved.

In the present work, following previous proposal\cite{koyama95}, we
seek continuously tunable emissions of THz electromagnetic waves
from high-$T_c$ superconductors with the principle of Josephson
relation\cite{Josephson}. For this purpose, the highly anisotropic
high-$T_c$ superconductor $\rm{Bi_2Sr_2CaCu_2O_{8+\delta}}$ can be
considered as a stack of Josephson junctions in atomic
scale\cite{kleiner92}. The novel device has its advantages since,
first, the energy gap in high-$T_c$ superconductors is much larger
than the plasma energy and thus the plasma, if excited in some way,
should be stable; secondly, the power output conjectured to be
proportional to the junction number squared would be very large;
thirdly, variations of physical parameters in these intrinsic
Josephson junctions (IJJs) are much smaller than artificially
fabricated junctions. For artificial Josephson junction arrays,
radiations of electromagnetric waves have been
demonstrated\cite{barbara99}; the frequency is, however, in the
sub-THz regime because of the small superconducting energy gap.

There are theoretical calculations\cite{bulaevskii06} as well as
numerical simulations\cite{machida01b,tac05} which discussed
possible radiation from IJJs of high-$T_c$ cuprates. However, many
issues have not yet been revealed concerning the mechanism of THz
emission. Although the Josephson plasma obviously play a key role
here, the physical parameters of single Josephson junctions of
$\rm{Bi_2Sr_2CaCu_2O_{8+\delta}}$ only give plasma frequency in
sub-THz regime. Therefore the collective modes in the stack of
Josephson junctions are essential in order to lift the frequency by
an order of magnitude. This goal is expected to be achieved by the
Josephson vortices driven by the $c$-axis current, which generates
oscillating voltage across junctions according to the ac Josephson
relation. The Josephson plasma and the motion of vortices intervene
with each other in a complex way, which makes the resonance
condition in this system obscure. Without revealing the resonance
mechanism, it is hard to tell whether a continuous tuning of
frequency is possible or not.

On the other hand, experimental efforts toward exciting THz
electromagnetic wave using IJJs seem to be accelerated recently
\cite{ustinov06,kadowaki06,bae07}. However, there are large
discrepancies in estimates of the optimal power output:
theoretically $3000\rm{W/cm^2}$ in Ref.~\cite{tac05},
$1500\rm{W/cm^2}$ in Ref.~\cite{bulaevskii06}, while experimentally
$150\rm{W/cm^2}$ in Ref.~\cite{kadowaki06}, $20\rm{W/cm^2}$ in
Ref.~\cite{bae07}, which hinders a clear assessment on the new
technique.

In this paper, we investigate the THz radiation from intrinsic
Josephson junctions by solving coupled nonlinear sine-Gordon
equations numerically using an appropriate boundary condition. The
main results of the present work are summarized as follows: Resonant
radiation of THz wave occurs at the edge of junctions due to the
cavity resonance of transverse plasma. A large resonance is achieved
when the velocity of Josephson vortices matches the velocity of
Josephson plasma. The vortex configuration is revealed to be
rectangular with additional random sliding motions at resonance. The
maximum energy is about $400\rm{W/cm^2}$ and the frequency covers
the THz band. Both the energy and frequency can be tuned almost
continuously by the bias current and applied magnetic field.

\vspace{3mm}

\begin{figure}[t]
\psfig{figure=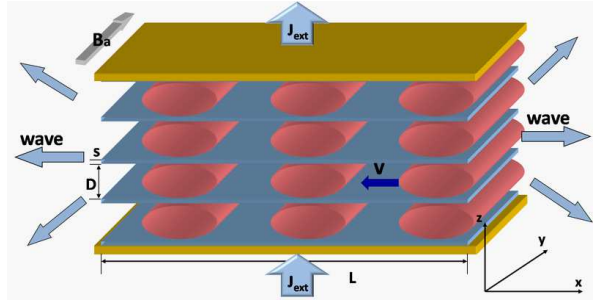,width=\columnwidth} \caption{\label{f1}(color
online). Schematic diagram of terahertz electromagnetic wave
emitter. The intrinsic Josephson junctions are sandwiched by two
gold electrodes through which a dc bias current is fed into the
system.}
\end{figure}

\noindent {\it Model equations --} The model we use, sketched
schematically in Fig. \ref{f1} is the same as Ref.\cite{tac05}. A
static magnetic field $B_a$ of order of $1$T is applied along the
$y$ axis, which induces Josephson vortices in all insulating layers.
 A bias current is fed
into the system along the $z$ axis, which drives fluxons towards the
negative direction of the $x$ axis. The moving fluxons excite the
transverse Josephson plasma\cite{tamasaku92,matsuda95}. For
simplicity, we ignore here the effect of thermal fluctuations and
thus the vortex lines are straight. This approximation makes one to
be able to reduce the three dimensional system to two dimensions.
While the estimate on the output energy should be considered as the
upper limit for experiments, the mechanism of radiation remains
unchanged even when thermal fluctuations are involved.

Using the London theory, Josephson relation as well as Maxwell
equations, the superconducting and electromagnetic properties of the
system can be described by the following
equations\cite{tac05,sakai93,machida99}:
\begin{equation}\label{eq1}
\begin{array}{l}
 (1 -\zeta\Delta ^{(2)})\{\partial_{t'}^2 P_{l+1,l}+\beta\partial_{t'} P_{l+1,l}+\sin
 P_{l+1,l} \\
 +\alpha s'[\partial _{t'}(\rho_{l+1}'-\rho_l')+\beta(\rho_{l + 1}'-\rho
 _l')]\}=\partial _{x'}^2 P_{l+1,l},
\end{array}
\end{equation}
\begin{equation}\label{eq2}
s'(1-\alpha \Delta ^{(2)})\rho _l'=\partial_{t'}(P_{l +
1,l}-P_{l,l-1}),
\end{equation}
where $P_{l+1,l}$ is the guage invariant phase difference defined as
\begin{equation}\label{eq3}
P_{l+1,l}(\mathbf{r},t)=\varphi_{l+1}(\mathbf{r},t)-\varphi_{l}(\mathbf{r},t)-\frac{2\pi}{\phi_0}\int_{z_l}^{z_{l+1}}A_z(\mathbf{r},t)dz,
\end{equation}
with the phase of order parameter $\varphi$, the vector potential
$A_z$ and the quantum flux $\phi_0$. The operator $\Delta^{(2)}$ is
defined as $\Delta^{(2)}f_l=f_{l+1}+f_{l-1}-2f_l$. In
Eqs.(\ref{eq1}) and (\ref{eq2}), dimensionless quantities are used:
$x'=x/\lambda_c$, $t'=t\omega_p$ and $\rho_l'=\rho_l\lambda_c
\omega_p/J_c$, where $\rho_l$ is the charge density in the $l$th
superconducting layer, $\omega_p=c/\lambda_c\sqrt{\varepsilon_c}$ is
the plasma frequency, $\varepsilon_c$ is the dielectric constant
along the $z$ axis. $\lambda_c$, and $\lambda_{ab}$ in below, are
the penetration depths. The other dimensionless quantities are
defined as: $J'=J/J_c$, $B'=2\pi\lambda_c D B/\phi_0$,
$\beta=4\pi\sigma_c\lambda_c/c\sqrt{\varepsilon_c}$,
$E'=\sigma_cE/\beta J_c$ , $\alpha=\varepsilon_c\mu^2/sD$
(capacitive coupling), $\zeta=\lambda_{ab}^2/sD$ (inductive
coupling) and $s'=s/\lambda_c$, where $\sigma_c$ is the
conductivity, $\mu$ is the Debye screening length, $J_c$ is the
critical current density, $s$ ($D$) is the thickness of the
superconducting (insulating) layer.

\vspace{3mm}

\noindent {\it Boundary condition --} Equations (\ref{eq1}) and
(\ref{eq2}) must be modified at the topmost and bottommost
junctions. We assume that the superconductivity will penetrate into
the gold electrodes\cite{tac05} so that the thickness of the topmost
and bottommost superconducting layers $s_0$ is larger than $s$. We
also assume the electrodes as good conductor, therefore the electric
field in the electrodes is zero. With these two assumptions, the
equations at the topmost and bottommost layer can be written down
straightforwardly\cite{koy96}.

The boundary condition at the edge ($x=0$ and $x=L$) is more subtle.
We implemented the dynamic boundary condition which is determined by
the electromagnetic wave in the dielectric medium coupled to the
junctions\cite{bulaevskii06,bulaevskii06prl}. Assuming that only
transverse waves are transmitted through the dielectric medium,
which becomes exact when the number of junctions is infinite, there
is a linear relation between the electric and magnetic fields. With
the Fresnel conditions, one arrives at following condition between
the scillating fields $\widetilde{B'}$ and $\widetilde{E'}$ on the
IJJs side of interface\cite{bulaevskii06prl}
\begin{equation}\label{eq5}
\widetilde{B}'^y  =  \mp \sqrt {\varepsilon' _d } \widetilde{E}'^z,
\end{equation}
where $+$ ($-$) means the wave propagating in the negative(positive)
$x$ direction, $\varepsilon' _d\equiv\varepsilon _d/\varepsilon_c$
is the dielectric constant of the dielectric medium $\varepsilon _d$
normalized by $\varepsilon _c$.  With the relations
$\partial_{x'}P_{l+1,l}={B'}_{l+1,l}^y$ and
$\partial_{t'}P_{l+1,l}={E'}_{l+1,l}^z$, we derive from
Eq.(\ref{eq5}) the dynamic boundary condition for the IJJs
\begin{equation}\label{eq6}
\partial_{x'}P_{l+1,l}-{B'}_a\pm\sqrt{\varepsilon'_d}(\partial_{t'}P_{l+1,l}-E'_{dc})=0,
\end{equation}
where the magnetic field induced by $J'_{\rm{ext}}$ is ignored
because it is negligibly small in comparison to $B'_a$. In high-bias
region $E'_{\rm{dc}}\approx J'_{\rm{ext}}/\beta$ (see \cite{tac05}).

The parameters used for simulations are $\lambda_{ab}=0.4\rm{\mu
m}$, $\lambda_c=200\rm{\mu m}$, $s=3\rm{{\AA}}$, $s_0=0.075\rm{\mu
m}$, $D=12\rm{\AA}$, $\mu=0.6\rm{\AA}$, $\alpha=0.1$, $\beta=0.02$
and $\varepsilon_d=\varepsilon_c=10$. The number of junctions is
$N=30$ and the length is $L=19.2\rm{\mu m}$. The applied magnetic
field is $B_a=1$T. The equations of motion Eqs.(\ref{eq1}) and
(\ref{eq2}) are integrated by the five-point Nordsieck-Gear
Predictor-corrector method. The time step in all the simulations is
set to $\Delta t'=0.0002$ and the mesh size is set to $\Delta
x'=0.0002$ ($0.1\lambda_{ab}$). We start to calculate physical
quantities when the system reaches steady states which are
identified by the criterion that the voltage only fluctuates weakly
around a fixed value.

\vspace{3mm}

\begin{figure}[t]
\psfig{figure=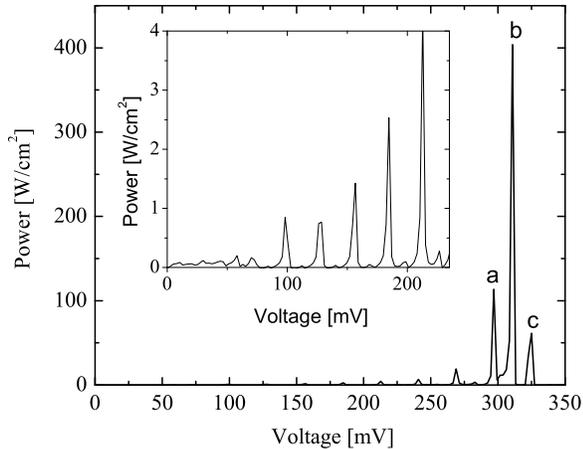,width=\columnwidth}
\caption{\label{f2}Radiation power counted by the Poynting vector as
a function of the voltage across the stack of IJJs. The external
current corresponds to the b-peak is $J_{\rm{ext}}=1.33J_{\rm{c}}$.
The inset is the enlarged view of the main panel in the region
$V<235$mV.}
\end{figure}

\noindent {\it Numerical results --}We gradually ramp up the current
$J'_{\rm{ext}}$ at $B_a=1$T and then gradually reduce it with step
$\Delta J'_{\rm{ext}}$=0.01. The output power is counted by the
Poynting vector at the left edge, toward which the vortices are
moving. The dependence of power on voltage $V$ is plotted in Fig.
\ref{f2}. The resonance takes place at discrete values of voltage
with equidistance between neighboring peaks.  The maximum power is
about $400\rm{W/cm^2}$.

\begin{figure}[b]
\psfig{figure=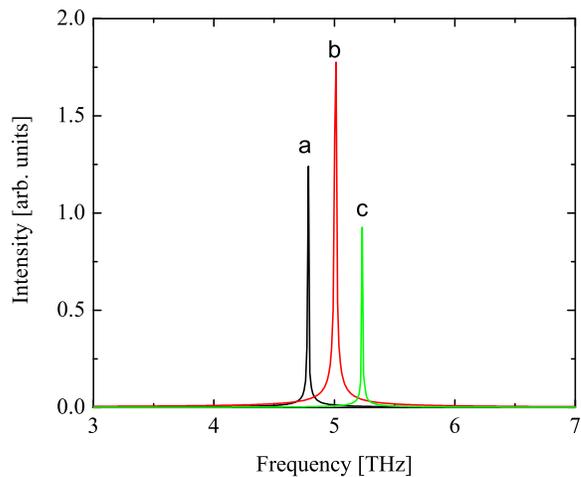,width=\columnwidth} \caption{\label{f3}(color
online). Frequency spectra for the voltage at the main peaks in
Fig.~\ref{f2}.}
\end{figure}

The spectra of the electric field at the main peaks in Fig.~\ref{f2}
obtained by fast Fourier transformation are displayed in
Fig.~\ref{f3}. The sharp spectra indicate monochromatic
electromagnetic waves in \emph{terahertz} regime. Thus we have
confirmed theoretically the terahertz laser radiation from the
intrinsic Josephson junctions.

The equidistance between neighboring peaks indicates a resonance
phenomenon, the one between Josephson plasma and cavity modes as
revealed below. The Josephson relation reads $\omega_{\rm
JP}=2eV/\hbar$ and the cavity modes read $\omega_{\rm c}=n\pi c_p/L$
where $c_p$ is the velocity of transverse plasma and $n$ is an
integer. The resonance occurs when $\omega_{\rm JP}=\omega_c$,
namely
\begin{equation}\label{eq7}
\overline{V}\equiv V/N=n\pi\hbar c_p/2eL.
\end{equation}
To be more specific, we counted the wave length $\lambda_w$,
frequency $f$ and the voltage $\overline{V}$ for the resonating
peaks in Fig.~\ref{f2}; a-peak: $\lambda_w=1.82\rm{\mu m}$,
$f=4.79$THz, $\overline{V}=9.90$mV; b-peak: $\lambda_w=1.74\rm{\mu
m}$, $f=5.01$THz, $\overline{V}=10.36$mV and c-peak:
$\lambda_w=1.66\rm{\mu m}$, $f=5.24$THz, $\overline{V}=10.83$mV. It
is easy to see that $f=2e\overline{V}/h$ and $n\lambda_w/2=L$ for
all the cases: a-peak, $n=21$; b-peak, $n=22$; c-peak, $n=23$. The
former is nothing but the Josephson relation and the latter is the
condition for cavity modes. From the wave length and the frequency,
we estimate the plasma velocity as $c_p\approx
8.71\times10^6\rm{m/s}$. It is then easy to see that the interval
between consecutive peaks $\Delta \overline{V}=\pi\hbar
c_p/2eL=0.47$mV is satisfied perfectly. Therefore our simulation
indicates clearly that the terahertz laser is caused by the cavity
resonance.

The above plasma velocities $c_p$ is very close to the largest
characteristic velocity $c_1=8.75\times10^6\rm{m/s}$ obtained by
solving the linearized Eqs.(\ref{eq1}) and (\ref{eq2}) with $s_0>s$
(see also \cite{sakai94}). The velocity $c_1$ associated with the
node-less mode is most important for resonance since it corresponds
to the in-phase motion of vortices.

\begin{figure}[t]
\psfig{figure=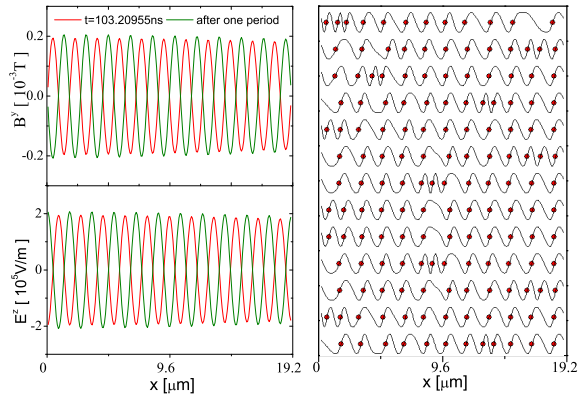,width=\columnwidth} \caption{\label{f6}(color
online). Left panel: snapshots for the magnetic and electric fields
at the largest resonance in Fig.~\ref{f2}, where the static
backgrounds have been subtracted. Right panel: snapshot of vortex
configuration.}
\end{figure}

From Fig.~\ref{f2}, it is clear that there is a bundle of large
resonances around $V=310.8$mV for $B_a=1$T. In order to reveal the
reason for this nonlinear property, we investigate the vortex
motion. Since $P_{l+1,l}=2\pi DB_ax/\phi_0+2\pi
cDE^z_{l+1,l}t/\phi_0 +\widetilde{P}_{l+1,l}$, where
$\widetilde{P}_{l+1,l}$ is the oscillating contribution, the
velocity of vortices is evaluated as $v\approx
cE^z_{l+1,l}/B_a=8.63\times10^6\rm{m/s}$ at the b-peak in
Fig.~\ref{f2}. It is close to the plasma velocity $c_p$ and $c_1$.
Therefore, the largest resonance takes place when the velocities of
vortices and transverse plasma coincide with each other. A similar
relation has been discussed in Ref.~\cite{koshelev00}, where an
infinitely long junction was considered and therefore no cavity mode
was involved.

\begin{figure}[b]
\psfig{figure=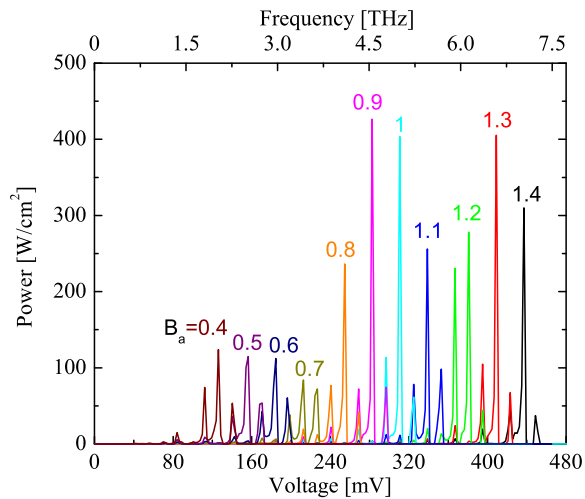,width=\columnwidth} \caption{\label{f5}(color
online). Radiation power at different applied magnetic fields in
units of Telsa as a function of the voltage across the stack of
IJJs. The frequency in the upper axis is determined by the ac
Josephson relation.}
\end{figure}

From $v=c_p=c_1$, it becomes clear that the largest energy emission
is excited by a voltage
\begin{equation}\label{eq8}
V=Nc_1 B_aD/c.
\end{equation}
With Eqs.~(\ref{eq7}) and (\ref{eq8}), we have $n=2B_aLD/\phi_0$.
Since $n\lambda_w/2=L$, it is concluded that the optimal output is
achieved when the wave length of plasma equals to the vortex-vortex
separation $\lambda_w=\phi_0/B_aD$.

 It is interesting to take a look at the
vortex configuration at resonance at this point. As seen from the
snapshot shown in Fig.~(\ref{f6}), the vortices form an overall
ordered rectangular lattice; sliding motions take place in random
places for the time being. The rectangular vortex lattice is in
accordance with the matching between the measured electromagnetic
wave velocity and the node-less plasma velocity.

Having clarified the mechanism of resonance, we turn to investigate
how to tune the resonance frequency and power. The voltage
dependence of power emission at different magnetic fields is
presented in Fig.~\ref{f5}. The voltage for largest power output
increases linearly with the magnetic field, consistent with
Eq.~\ref{eq8}. From the ac Josephson relation, as depicted in the
upper axis of Fig.~\ref{f5}, the frequency of emitted
electromagnetic wave can be tuned by sweeping the magnetic field and
adjusting the dc voltage accordingly. Although the cavity resonance
imposes that the resonating frequency is discrete when voltage is
swept, it becomes almost continuous if the length of IJJs $L$ is
large enough. For example, the frequency interval between
neighboring resonances is about $0.04\rm{THz}$ for $L=100\rm{\mu
m}$.

At a given voltage, on the other hand, the emitted power shrinks
quickly when the magnetic field is tuned away from the optimal
matching value. This permits one to control the power output in a
sensitive way.

\vspace{3mm}

\noindent {\it Discussions --} Finally we compare briefly our
results with recent experimental works. Both Refs.~\cite{kadowaki06}
and \cite{bae07} tried to stimulate THz electromagnetic waves
utilizing the Josephson vortex dynamics. While the emission was
detected by bolometer in Ref.~\cite{kadowaki06}, Ref.~\cite{bae07}
used a second stack of IJJs to detect excited THz electromagnetic
waves. It is somewhat difficult to draw conclusions on the resonance
mechanism from experiments; in our simulation, however, cavity
resonances are clearly observed. In Ref.~\cite{bae07} it is found
that the most efficient emission is achieved for the rectangular
vortex configuration, same as our computer simulation. Although not
revealing a clear relation, both works tried to tune the frequency
of emission by sweeping simultaneously the voltage and magnetic
field, which, according to our theoretical investigation, is an
efficient way for achieving the frequency tunability. In our
simulation, we use magnetic fields below $1.5$ Tesla, which are
quite smaller than those in Ref.~\cite{bae07}. According to our
relation for the optimal matching magnetic field and voltage, in
order to get large emission power under $4$ Tesla the voltage should
be unrealistically large. It can be a possible reason that our
theoretical estimate on the power output is different from that in
the experiment.

\vspace{3mm}

\noindent {\it Acknowledgements --} X.H. acknowledges Q.-H. Chen and
Y. Nonomura for discussions at the early stage of this study.
Calculations were performed on SR11000 (HITACHI) in NIMS. X. H. is
supported by Grant-in-Aid for Scientific Research (C) No. 18540360
and CTC program of JSPS, and project ITSNEM of China Academy of
Science.


\end{document}